\documentclass[preprint,12pt]{revtex4-1}
\usepackage{graphicx}
\usepackage{epsfig}
\usepackage{color}
\usepackage{epstopdf}

\usepackage{amssymb}

\def\be{\begin{equation}}
\def\bea{\begin{eqnarray}}
\def\ee{\end{equation}}
\def\eea{\end{eqnarray}}

\def\pt{\partial}

\def\Dt{\Delta}
\def\eps{\varepsilon}
\def\ffi{\varphi}

\def\om{\omega}

\def\dd{\mbox{d}}
\def\sign{\mbox{sign}}

\def\const{\mbox{const}}

\begin{document}


\title{Kinetic equation for systems with resonant captures and scatterings.} 



\author{A. V. Artemyev}
 \altaffiliation[Also at ]{Space Research Institute, RAS, Moscow, Russia}
\email{aartemyev@igpp.ucla.edu}
\affiliation{
Institute of Geophysics and Planetary Physics, UCLA, Los Angeles, California, USA.
}%

\author{A. I. Neishtadt}
 \altaffiliation[Also at ]{Space Research Institute, RAS, Moscow, Russia}
\affiliation{
Department of Mathematical Sciences, Loughborough University, Loughborough LE11 3TU, UK.
}%

\author{A. A. Vasiliev}
\affiliation{Space Research Institute, RAS, Moscow, Russia.}

\author{D. Mourenas}
\affiliation{CEA, DAM, DIF, Arpajon, France}

\date{\today}

\begin{abstract}
We study a Hamiltonian system of type describing a charged particle resonant interaction with  an electromagnetic wave. We consider an ensemble of particles that repeatedly pass through the resonance with the wave, and study evolution of the distribution function due to multiple scatterings on the resonance and trappings (captures) into the resonance. We derive the corresponding kinetic equation.   Particular cases of this problem has been studied in our recent papers \cite{ANVM16,ANVM17}.

\end{abstract}


\maketitle 

\section{Introduction}

Resonant phenomena are a key part in long-term evolution of numerous systems in plasma physics, hydrodynamics, celestial mechanics, etc. The phenomena of scattering on a resonance and capture (trapping) into a resonance were described in details in \cite{Neishtadt75,Neishtadt99} (see also \cite{bookAKN06,NV06}), and all the characteristics of a single passage through a resonance were obtained. These results were applied to  studies of the resonant phenomena in various problems in physics; among recent studies  we just mention papers \cite{Itin00,Vainchtein04:prl,Neishtadt11:mmj,Vasiliev11,Artemyev10:chaos,Artemyev15:pre}. However, in physical systems one has usually to deal with an ensemble of particles (phase trajectories), which pass repeatedly through the resonance during long time intervals. These multiple resonant interactions affect  the distribution function of the ensemble. Thus a crucial issue is to implement the properties of individual resonant interactions into a kinetic description of evolution of the distribution function.

A major peculiarity on this way is that captures into  resonances provide fast and large-distance transport in the phase space, which cannot be described with differential operators in the kinetic equation. In the papers \cite{Shklyar81,Artemyev14:grl:fast_transport,Omura15}, it was proposed to introduce integral operators describing this kind of transport. This approach, however, did not take into account  kinetic balance between the captures and the scattering. Namely, while rare captures result in strong variation (say, growth) of energy of a small part of particles (phase trajectories), scatterings produce small energy variation in the opposite direction (decrease) of a large sub-ensemble. Therefore, to include these phenomena into the kinetic equation, one should find and implement the relationship between the corresponding kinetic coefficients. This approach was first proposed in \cite{ANVM16} in the simplest case of a Hamiltonian system with one and a half d.o.f., and in \cite{ANVM17} for a more realistic system with two d.o.f. In these papers, we have introduced a Fokker-Planck kinetic equation describing evolution of an ensemble of particles in a system where repeated scatterings on resonances and captures into resonances (followed by escapes from the resonances) take place. Our approach is based on the fact that one can introduce probability of capture into a resonance, and that this probability turns out to be interconnected with the velocity of the drift in the phase space due to scatterings on the resonance.

In the present work we derive the kinetic equation in a general case when the time period between successive passages through the resonance depends on the particle energy.  In Section 2, we briefly outline the main approaches and results concerning an individual resonance crossing. In Section 3, we use these results to construct the kinetic equation describing the long-term evolution of the distribution function in a system with multiple resonant captures and scatterings. Note that in \cite{ANVM17} the similar equation  was obtained with  smaller terms omitted. In the present paper, these terms are taken into account allowing to represent the kinetic equation in a more elegant form.

\section{Resonant phenomena in slow-fast Hamiltonian systems}

Consider a  Hamiltonian system with Hamiltonian
\be
H = H_0(p,q) + \eps A(p,q)\sin(kq-\om t),
\label{2.1}
\ee
where $\eps$ is a small parameter and  $(p,q)$ are canonically conjugate variables. Such Hamiltonians naturally appear in problems of motion of a charged particle in a harmonic electromagnetic wave and a background magnetic field. This is a Hamiltonian system with 1$\frac12$ degrees of freedom. Introduce $t$ as a new canonical coordinate $u$, and $U$ as the canonically conjugate momentum. The Hamiltonian takes the form
$$
H = U + H_0(p,q) + \eps A(p,q)\sin(kq-\om u).
$$
Now we introduce the phase of the wave as an independent variable $\ffi = q- \om u/k$. To do this, we make a canonical transformation $(p,q,U,u) \mapsto (\hat p, \hat q, I, \ffi)$ using generating function
$$
W = I(q-\frac{\om}{k}u) + \hat p q + \hat U u
$$
Omitting constant $\hat U$ and omitting hats over $p$ and $q$, we obtain a 2 degrees of freedom Hamiltonian (we keep the same notations for the functions $H_0$ and $A$):
\be
H = -\frac{\om}{k}I + H_0(p,q,I) + \eps A(p,q,I)\sin (k\ffi)
\label{2.2}
\ee
Now we rescale the variables introducing $\bar \ffi = k\ffi$. In order to keep the symplectic structure, we also rescale time introducing $\bar t = kt$ and consider $(p,kq)$ as a pair of canonically conjugate variables. We assume that $k^{-1} = \eps$. Omitting the bars we obtain the Hamiltonian in the new variables:
\be
H = H_0(p,q,I) - v_{\phi}I +\eps A(p,q,I) \sin\ffi,
\label{2.3}
\ee
where we have used the notation $v_{\phi} = \om/k$. One can see from (\ref{2.3}) that with the accuracy of order $\sim\eps$ the  system stays on the energy level  $H_0(p,q,I) - v_{\phi}I =\const$; thus, we obtain the following relation between the particle energy $h = H_0$ and the value of $I$:
\be
h - v_{\phi}I = \const .
\label{2.40}
\ee

In Hamiltonian  (\ref{2.3}), the pairs of conjugate variables are $(p,\eps^{-1}q)$ and $(I,\ffi)$.
The equations of motion in the main approximation are
\bea
\dot p &=& -\eps\frac{\pt H_0}{\pt q}
\nonumber \\
\dot q &=& \eps\frac{\pt H_0}{\pt p}
\label{2.30} \\
\dot I &=& -\eps A \cos\ffi
\nonumber \\
\dot \ffi &=&\frac{\pt H_0}{\pt I} -v_{\phi} .
\nonumber
\eea
Thus, in this system variable $\ffi$ is a fast phase, and the other variables are slow.  Far from the resonance $\dot \ffi = 0$  the equations of motion can be averaged over the fast phase. Thus we obtain the averaged system:
\be
\dot p = -\eps\frac{\pt H_0}{\pt q}, \quad \dot q = \eps\frac{\pt H_0}{\pt p}, \quad \dot I = 0
\label{2.4}
\ee
Variable $I$ is the integral of the averaged system (\ref{2.4}) and hence is an adiabatic invariant of the exact system (\ref{2.3}) (see, e.g., \cite{bookAKN06}). Far from the resonance, it is preserved with a good accuracy along phase trajectories of (\ref{2.3}).

We assume that the slow motion on the $(p,q)$-plane in the averaged system (\ref{2.4}) is periodic. The area bounded by a trajectory of this averaged motion can be considered as a function of the energy $H_0=h$ or of the corresponding value of $I$ (see (\ref{2.40})).  The condition of resonance ${\pt H_0}/{\pt I}=0$ defines a curve on the $(p,q)$-plane (the resonant curve). In a general situation, trajectories of the averaged system cross the resonant curve.

In a small vicinity of the resonance, the averaging of equations (\ref{2.30}) does not work properly, and here we apply the standard approach developed in \cite{Neishtadt99} (see also, e.g., \cite{bookAKN06,NV06}). We expand the Hamiltonian $H$ into series near the resonant value of $I=I_R$, where $I_R = I_R(p,q)$ is found from the equation ${\pt H_0}/{\pt I}=v_{\phi}$. Thus we obtain Hamiltonian
\be
H = \Lambda(p,q) + \frac12 g(p,q)(I-I_R)^2 + \eps A(p,q,I_R)\sin\ffi,
\label{2.5}
\ee
where $\Lambda = H_0(p,q,I_R)$ and $g = \left.\pt^2 H_0/\pt I^2 \right|_{I=I_R}$ and smaller terms are omitted. Introduce new canonical momentum $K = I-I_R$ with the generating function $W=\bar p \eps^{-1}q + (K+I_R)\ffi$,  where $(\bar p,\bar q)$ are new variables. In the new variables the Hamiltonian takes the form (bars are omitted, we keep the same notations for the functions $\Lambda$ and $A$):
\be
H = \Lambda(p,q) + \frac12 g(p,q)K^2 + \eps A(p,q)\sin\ffi +\eps\beta(p,q)\ffi \equiv \Lambda(p,q) + F,
\label{2.6}
\ee
where $\beta(p,q) = \{I_R,\Lambda\}$,  $\{\cdot,\cdot\}$ denotes the Poisson bracket with respect to $(p,q)$, and we have introduced the so-called pendulum-like Hamiltonian $F$. The coefficients $g,A,$ and $\beta$ in $F$ depend on slow variables $p,q$, while the evolution of $p,q$ is defined by Hamiltonian $\Lambda$. If $A(p,q)>\beta(p,q)$, the phase portrait of $F$ on the $(\ffi,K)$-plane has a saddle point and a separatrix, see Fig. 1. The area $S$ of the region inside the separatrix loop can be found as
\be
S(p,q)=2 \int_{\ffi_{min}}^{\ffi_1} K \dd\ffi = \sqrt\eps \int_{\ffi_{min}}^{\ffi_1} \sqrt{\frac{2}{g}\left(\frac{F_s}{\eps} - A\sin\ffi - \beta\ffi \right)} \dd\ffi,
\label{2.7}
\ee
where $F_s$ is the value of $F$ at the saddle point, $\ffi_1$ and $\ffi_{min}$ are shown in Fig. 1.

\begin{figure}
\begin{centering}
\includegraphics[width=14cm]{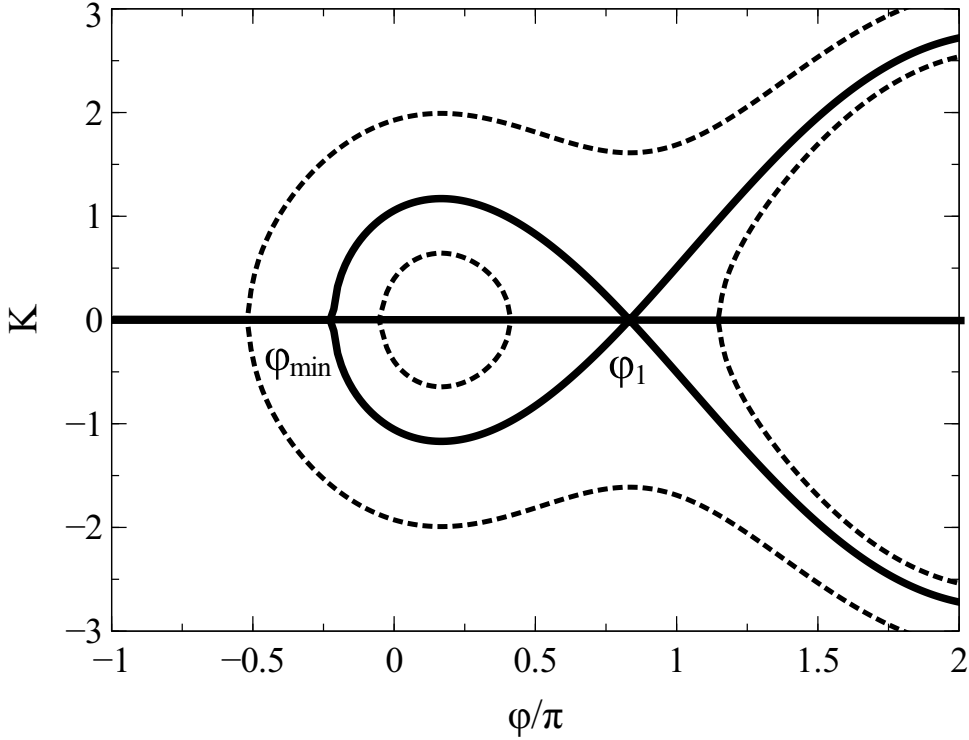}
\par\end{centering}
\caption{Phase portrait of Hamiltonian $F$ in (\ref{2.6}) in the case $A(p,q)>\beta(p,q)$. It is assumed that $g,\beta$ are positive. \label{fig:phase}}
\end{figure}

Closed phase trajectories on the phase portrait of the pendulum-like Hamiltonian $F$ correspond to phase points captured into the resonance, while open trajectories correspond to those passing through the resonance.  If $S$ grows, there appears additional phase volume inside of the separatrix loop, and phase points can be captured into the resonance. Motion on the phase portrait is fast compared to the speed of variation of $p,q$. Hence, the area surrounded by a captured trajectory is an adiabatic invariant of this system. Therefore, while the area $S$ grows, the phase point stays within the separatrix loop. If later $S$ decreases, the phase point can leave the separatrix loop when the area $S$  again equals the same value as at the time of capture. This is an escape from the resonance. Hence to predict the escape from the resonance one can use the time profile of the function $S(p,q) = S(t)$ along the resonant trajectory where the evolution of $(p,q)$ is defined by the Hamiltonian $\Lambda$. On the other hand, capture into the resonance is possible only if the phase point approaches the resonance when the function $S(t)$ grows.

While a phase point is captured, the corresponding value $h$ of the Hamiltonian $H_0$ of the averaged system (\ref{2.4}) varies with time. The value $h$ can be used to parametrize function $S$, and it is useful to consider $S$ as a function of $h$: $S=S(h)$. We assume that $S(h)$ has the only maximum at $h = h_{max}$ (see Fig. 2). Thus, phase points captured at $h_- < h_{max}$ are transported in Fig. 2 to  the right and escape from the resonance at $h_+ < h_{max}$ such that $S(h_+)=S(h_-)$. One can see that a capture followed by escape from the resonance result in strong (of order $1$) variation of the value of $h$ (and of the value of $I$, see (\ref{2.40})).

\begin{figure}
\begin{centering}
\includegraphics[width=14cm]{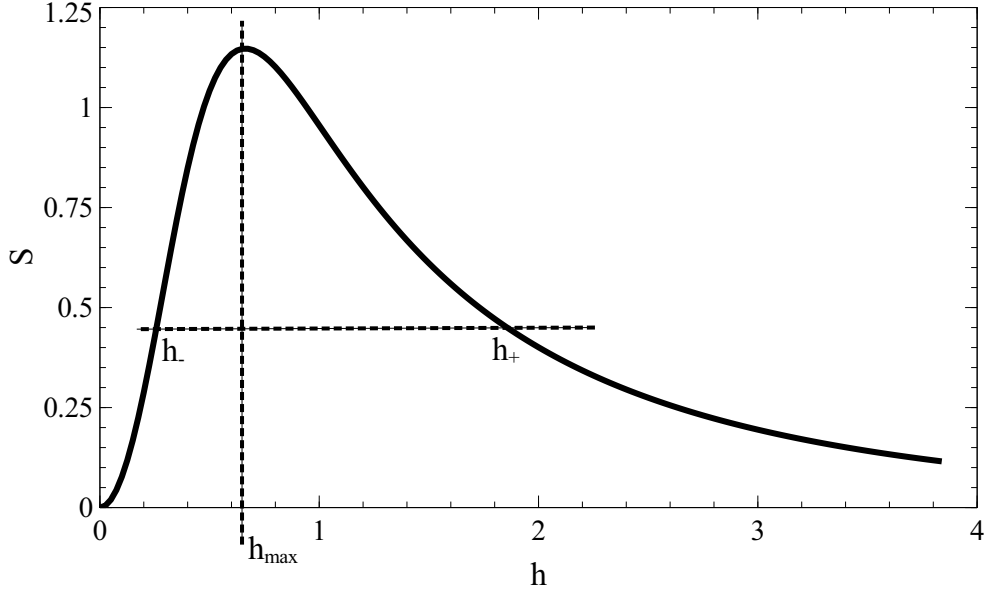}
\par\end{centering}
\caption{Plot of the area $S$ as a function of the particle energy $h$. \label{fig:plot}}
\end{figure}

Capture into a resonance is a probabilistic process (see \cite{Neishtadt99}). Consider a small time interval $\Dt t$.  The probability of capture can be calculated as the ratio of the number of phase points captured into the resonance during this interval (i.e., $\sim \Dt t \dot S$) to the total number of phase points crossing the resonant curve. Thus one obtains the following formula for the probability of capture into the resonance:
\bea
\Pi = \frac{\{S,\Lambda\}}{2\pi |\beta|}, \,\,\,\, \mbox{if}\,\,\, \{S,\Lambda\}>0,\nonumber\\
\label{2.8}\\
\Pi = 0,\,\,\,\, \mbox{if}\,\,\, \{S,\Lambda\} \le 0.
\nonumber
\eea
One can see from (\ref{2.8}) and (\ref{2.7}) that the capture probability is a small value of order $\sqrt\eps$.

Phase points that cross the resonant curve without capture are scattered on the resonance. The scattering results in a small variation $\Dt I \sim \sqrt\eps$. Exact amplitude of scattering is a random value (see, e.g., \cite{Neishtadt99,NV06}). If we have an ensemble of phase points, the mean scattering amplitude is (see \cite{Neishtadt99})
\be
\langle \Dt I \rangle = -\sign (\beta)\frac{S}{2\pi},
\label{2.9}
\ee
where $\langle \cdot \rangle$ denotes the ensemble average.
In terms of the particle energy $h$, the mean scattering amplitude is
\be
\langle \Dt h \rangle = -\sign (\beta)\frac{S}{2\pi} v_{\phi}.
\label{2.10}
\ee

To summarize, suppose we have an ensemble of phase points with the same initial value of $h$. After crossing the resonance, a small part of this ensemble given by (\ref{2.8}) is captured into the resonance and its energy significantly changes. The other phase points of the original ensemble are scattered on the resonance with the mean variation of energy given by (\ref{2.10}). Generally speaking, on each period $\tau(h)$ of the slow motion  a phase trajectory of the averaged system crosses the resonance several times.  Assume for simplicity that  $A \ne 0$ at only one of these crossings.  (Such situations occur in physical problems, see, e.g., \cite{ANVM17}).   Repeated passages through the resonance result  in drift and diffusion of $h$. Introduce the drift velocity  and the diffusion coefficient as
\be
V_h =\langle \Dt h \rangle /\tau(h), \,\,\, D_{hh} = \langle (\Dt h)^2 \rangle /\tau(h).
\label{eq2}
\ee

Next step is to establish the relation between the capture probability $\Pi$ and the drift velocity $V_h$. From (\ref{2.8}) and (\ref{2.40})  one obtains (if $\{S,\Lambda\}>0$):
\be
\Pi = \frac{1}{2\pi |\beta|} \frac{\dd S}{\dd h}\frac{\dd h}{\dd I}\left. \frac{\dd I}{\dd t}\right|_{I=I_R}= \frac{1}{2\pi |\beta|} \frac{\dd S}{\dd h} v_{\phi} \beta =  \frac{v_{\phi}}{2\pi} \sign(\beta)\frac{\dd S}{\dd h}.
\label{2.11}
\ee
Comparing this expression with (\ref{2.10}), we find
\be
\Pi = - \frac{\dd \langle \Dt h \rangle}{\dd h}.
\label{2.12}
\ee

\section{Evolution of the distribution function.}

Consider the distribution function of the phase points $f(h,t)$. The kinetic equation for this distribution function has a general form
\be
\frac{\pt f}{\pt t} = L_s f+ L_c f,
\label{eq0}
\ee
where operators $L_s$ and $L_c$ are related to scattering and capture/escape processes, respectively. The scattering part has a standard form
\be
L_s f = - \frac{{\partial (f V_h)}}{{\partial h}} + \frac12 \frac{{\partial }}{{\partial h}}\left( {D_{hh} \frac{{\partial f}}{{\partial h}}} \right) + L_{sm} f.
\label{eq1}
\ee
Here $V_h, D_{hh}$ are drift and diffusion coefficients respectively, defined in the previous section, and $L_{sm}$ is an additional small ($\sim D_{hh}$) drift term. This term appears because $V_h$ is calculated in the principal order in $\sqrt\eps$, and it will be omitted in the following consideration.

We  assume that  the function $S(h)$ has only  one maximum at $h = h_{max}$. The capture/escape operator in (\ref{eq0}) has different forms for $h<h_{max}$ (capture) and $h>h_{max}$ (escape from the resonance). In the case of capture, $h<h_{max}$, we have
\be
L_c f = -\frac{\Pi(h) f}{\tau},
\label{eq3}
\ee
where $\Pi(h)$ is the probability of capture and $\tau = \tau(h)$ is the period of the averaged motion. Using (\ref{2.12}) we find from (\ref{eq3})
\be
L_c f = \frac{f}{\tau}\frac{\dd \langle \Dt h \rangle}{\dd h}.
\label{eq4}
\ee

In the case of escape, $h>h_{max}$, introduce $h_*$ as the value of the energy that the phase point had before the capture to escape with energy $h$. Denote $\Pi_* = \Pi(h_*), \, \tau_* = \tau(h_*), \, f_* = f(h_*,t)$. Then we have
\bea
L_c f &=& \frac{\Pi_* f_*}{\tau_*}\left|\frac{\dd h_*}{\dd h} \right| = -\frac{\Pi_* f_*}{\tau_*}\frac{\dd h_*}{\dd h} =  -\frac{\Pi_*}{\tau_*}\frac{\dd S(h)/\dd h}{\dd S(h_*)/\dd h_*} f_* \nonumber \\
&=& - \frac{v_{\phi}}{\tau_*} \sign(\beta) \frac{\dd S(h_*)/\dd h_*}{2\pi }\frac{\dd S(h)/\dd h}{\dd S(h_*)/\dd h_*} f_* \nonumber \\
&=& - \frac{v_{\phi}}{\tau_*} \sign(\beta) \frac{\dd S(h)/\dd h}{2\pi } f_* = \frac{\dd \langle \Dt h \rangle}{\dd h} \frac{f_*}{\tau_*}. \nonumber
\eea
Substituting the above expressions into (\ref{eq0}) and using (\ref{eq2}) we  obtain the following form of the kinetic equation:

At $h<h_{max}$
\be
\frac{{\partial f}}{{\partial t}} = - V_h \frac{{\partial f}}{{\partial h}} + \frac{1}{\tau}\frac{\pt\tau}{\pt h} V_h f + \frac12 \frac{{\partial }}{{\partial h}}\left( {D_{hh} \frac{{\partial f}}{{\partial h}}} \right)  ;
\label{eq6}
\ee

at $h>h_{max}$
\be
\frac{{\partial f}}{{\partial t}} = - V_h \frac{{\partial f}}{{\partial h}} - \frac{{\pt V_h }}{{\pt h}}\left( {f - f_*\frac{{\tau}}{{\tau_*}} } \right) + \frac{1}{\tau_*}\frac{\pt\tau}{\pt h} V_h f_* + \frac12 \frac{{\partial }}{{\partial h}}\left( {D_{hh} \frac{{\partial f}}{{\partial h}}} \right) .
\label{eq7}
\ee
In \cite{ANVM17}, we omitted smaller terms with $\tau^{-1}\pt\tau/\pt h$ in equation (\ref{eq6})-(\ref{eq7}). This does not affect significantly the numerical results. However, now we keep these terms to proceed to a more concise form of the kinetic equation.

One can rewrite kinetic equation (\ref{eq6})-(\ref{eq7}) using the action variable of the averaged system $J$ instead of the energy $h$. According to the Hamiltonian equations of motion, these two variables are interconnected via $\pt h/\pt J = 2\pi/\tau$. Using this relation we introduce $\tilde f(J,t), \, V_J, \, D_{JJ}$ in place of $f(h,t), \, V_h, \, D_{hh}$ in the kinetic equation and take into account that
\be
f = \frac{\tilde f \tau}{2\pi}, \,\,\, V_h = \frac{2\pi V_J}{\tau}, \,\,\, D_{hh} = \frac{4\pi^2 D_{JJ}}{\tau^2}.
\ee
After straightforward calculations we finally obtain the kinetic equation in terms of the action $J$ (we omitted tildes over $f$):

At $h<h_{max}$
\be
\frac{{\partial f}}{{\partial t}} = - V_J \frac{{\partial f}}{{\partial J}} + \frac12 \frac{{\partial }}{{\partial J}}\left( {D_{JJ} \frac{{\partial f}}{{\partial J}}} \right)  ;
\label{eq8}
\ee

at $h>h_{max}$
\be
\frac{{\partial f}}{{\partial t}} = - V_J \frac{{\partial f}}{{\partial J}} - \frac{{\pt V_J }}{{\pt J}}\left( f - f_*  \right) +  \frac12 \frac{{\partial }}{{\partial J}}\left( {D_{JJ} \frac{{\partial f}}{{\partial J}}} \right) .
\label{eq9}
\ee

One can find numerical evidence supporting validity of kinetic equations (\ref{eq6}-\ref{eq9}) in our paper \cite{ANVM17}.

\section*{Acknowledgements}

The work of A. Artemyev, A. Neishtadt, and A. Vasiliev   was supported by
the Russian Scientific Fund, Project No. 14-12-00824.

\end{document}